\documentclass{technote}

\graphicspath{{./images/}}

\title{Equivalent Mechanical Models for Sloshing}
\authorsInShort{F. Capolupo}
\author[1]{Francesco Capolupo}
\affil[1]{European Space Agency, 2201 AZ Noordwijk, The Netherlands}

\makeglossaries

\begin{document}

\startdocshort


%

\section*{Nomenclature}
\begin{tabular}{lll}
$\mathcal{R} = \left(O, {\bm{x}}_\mathcal{R}, {\bm{y}}_\mathcal{R}, {\bm{z}}_\mathcal{R}\right)$ & Reference frame $\mathcal{R}$, centered in $O$, and with axes ${\bm{x}}_\mathcal{R}, {\bm{y}}_\mathcal{R}, {\bm{z}}_\mathcal{R}$\\
${\bm{r}}_{AB}^\mathcal{C}$ & Position vector of point $A$ with respect to point $B$, projected in frame $\mathcal{C}$\\
$R_{\mathcal{A}\mathcal{B}}$ & Transformation matrix from frame $\mathcal{B}$ to frame $\mathcal{A}$, such that ${\bm{v}}^\mathcal{A} = R_{\mathcal{A}\mathcal{B}}{\bm{v}}^\mathcal{B}$\\
${\bm{\omega}}_{\mathcal{A}\mathcal{B}}^\mathcal{C}$  & Angular rate of $\mathcal{B}$ with respect to $\mathcal{A}$, projected in frame $\mathcal{C}$\\
$\crossmat{{\bm{v}}}$ & Cross product matrix of vector ${\bm{v}}$, such that $\crossmat{{\bm{v}}}{\bm{u}} = {\bm{v}}\times {\bm{u}}$\\
${\bm{a}}\otimes{\bm{b}}$ & Dyadic product of vectors ${\bm{a}}$ and $\bm{b}$, such that ${\bm{a}}\otimes{\bm{b}} = {\bm{a}}{\bm{b}}^\top$\\
${\bm{e}}_x = \left(1, 0, 0\right)$ & Unitary (column) vector along $x$\\
${\bm{e}}_y = \left(0, 1, 0\right)$ & Unitary (column) vector along $y$\\
${\bm{e}}_z = \left(0, 0, 1\right)$ & Unitary (column) vector along $z$
\end{tabular}

\section{Introduction}
Propellant sloshing is a well-known, but not completely mastered phenomenon in space vehicles. It is particularly critical in both microgravity environments—such as interplanetary spacecraft requiring high pointing stability—and high-g conditions, as encountered during launch, re-entry, and landing. In both cases, sloshing can significantly affect vehicle performance and stability, and must often be explicitly considered in the design of the guidance, navigation, and control (GNC) subsystem.

For stability analysis and control design, the most common approach to modeling sloshing is through an equivalent mechanical representation \cite{b1}, where the moving propellant is treated as a mechanical system interacting with the rigid (or flexible) spacecraft. Pendulum-based models and mass-spring-damper systems are widely used by control analysts to assess sloshing-induced perturbations on vehicles subjected to persistent non-gravitational acceleration along one of their body axes.

In this work, we present a rigorous mathematical formulation of pendulum dynamics, starting from a single spherical pendulum attached to a rigid spacecraft. We derive the nonlinear equations of motion for this 8-degree-of-freedom multi-body system, and then extend the formulation to include multiple pendulums, representing multiple sloshing modes within a tank and/or multiple tanks on the same vehicle. Furthermore, we derive the corresponding linearized equations of motion, explicitly accounting for a nominal longitudinal force acting on the vehicle -- consistent with the high-g sloshing regime -- expressed in either the inertial or body frame. Finally, we demonstrate the mathematical equivalence between the pendulum and mass-spring-damper models and validate the proposed models through time-domain simulation and frequency-domain analysis.

\section{Nonlinear pendulum model}
\subsection{Single pendulum}

\begin{figure}
    \centering
    \includegraphics[scale=0.60]{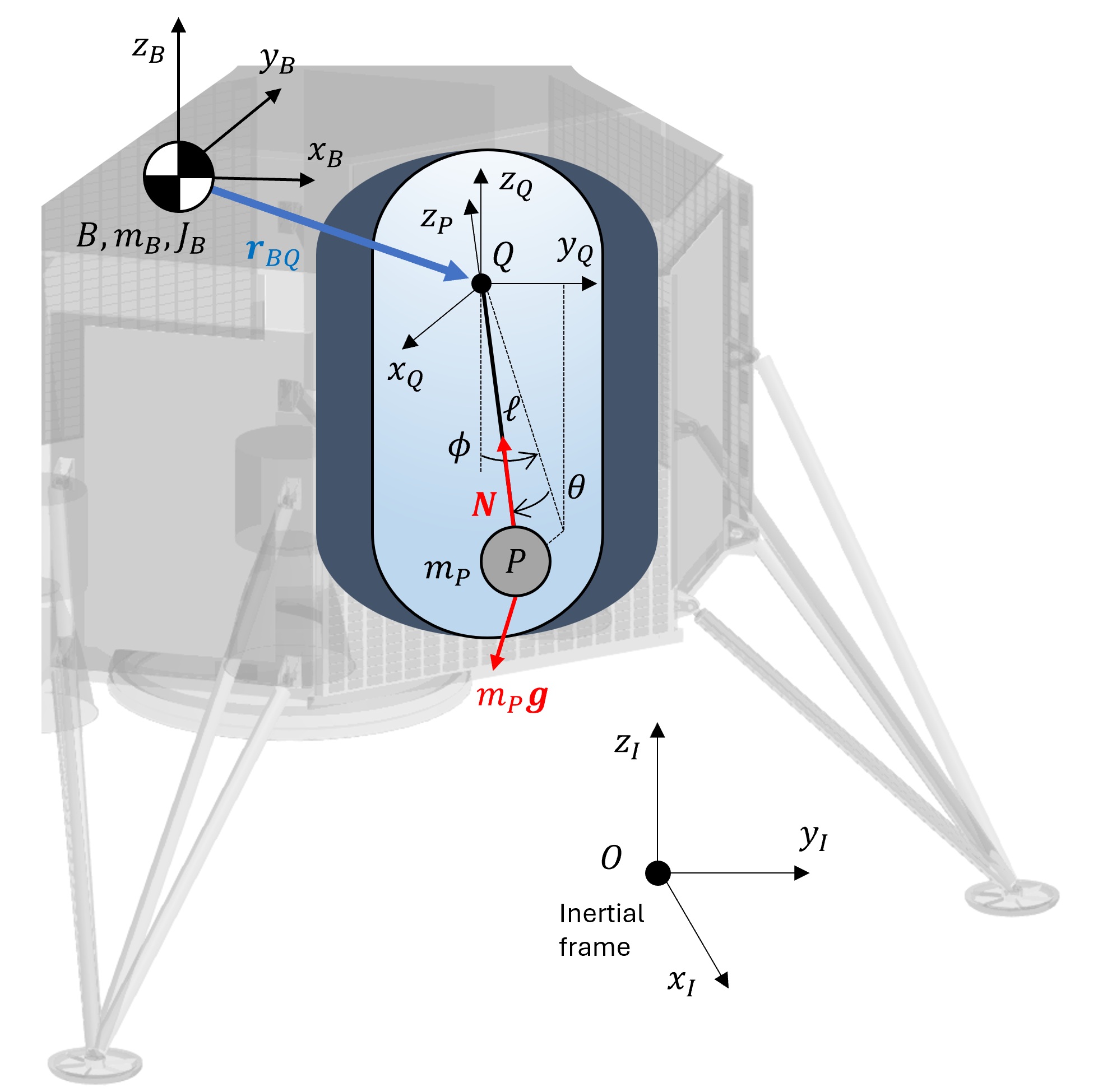}
    \caption{Spherical pendulum in a rigid body}
    \label{fig:pendul2}
\end{figure}

We consider a spherical pendulum $P$ of mass $m_P$ and constant length $\ell$, attached to a fulcrum $Q$ rigidly connected to a rigid body $B$ of mass $m_B$, center of mass $B$, and inertia tensor $J_B$, as illustrated in Figure \ref{fig:pendul2}. Together, the pendulum and the rigid body form a multi-body system with eight degrees of freedom. To describe its motion, we introduce an inertial reference frame $\mathcal{I} = \left(O, {\bm{x}}_\mathcal{I}, {\bm{y}}_\mathcal{I}, {\bm{z}}_\mathcal{I}\right)$, a rigid body-fixed frame $\mathcal{B} = \left(B, {\bm{x}}_\mathcal{B}, {\bm{y}}_\mathcal{B}, {\bm{z}}_\mathcal{B}\right)$, a fulcrum frame $\mathcal{Q} = \left(Q, {\bm{x}}_\mathcal{Q}, {\bm{y}}_\mathcal{Q}, {\bm{z}}_\mathcal{Q}\right)$, whose position and orientation are fixed with respect to $\mathcal{B}$, and a pendulum frame $\mathcal{P} = \left(Q, {\bm{x}}_\mathcal{P}, {\bm{y}}_\mathcal{P}, {\bm{z}}_\mathcal{P}\right)$, defined such that ${\bm{r}}_{QP} = -\ell {\bm{z}}_\mathcal{P}$. The orientation of $\mathcal{P}$ relative to $\mathcal{Q}$ is parametrized by two angles: a rotation of $\phi$ about ${\bm{x}}_\mathcal{Q}$, followed by a rotation of $\theta$ about $-{\bm{y}}_\mathcal{P}$. The corresponding transformation matrix from $\mathcal{P}$ to $\mathcal{Q}$ is therefore given by
\begin{equation}
    R_{\mathcal{Q}\mathcal{P}} = \mymatrix{\cos \theta & 0 & -\sin\theta\\
    -\sin\theta \sin \phi & \cos \phi & -\cos\theta \sin \phi\\
    \sin\theta \cos \phi & \sin \phi & \cos \theta\cos \phi}
\end{equation}
Consequently, the position of $P$ in the fulcrum frame is given by
\begin{equation}
    {\bm{r}}_{QP}^\mathcal{Q} = \ell\mymatrix{\sin\theta\\\cos\theta\sin\phi\\-\cos\theta\cos\phi}\label{defr}
\end{equation}
whereas the angular rate and angular acceleration of $\mathcal{P}$ with respect to $\mathcal{Q}$ are equal to
\begin{equation}
    {\bm{\omega}}^\mathcal{P}_{\mathcal{Q}\mathcal{P}} = \dot\phi {\bm{x}}_\mathcal{Q}^\mathcal{P} - \dot \theta {\bm{y}}_\mathcal{P}^\mathcal{P} = \mymatrix{\dot\phi\cos\theta \\ -\dot \theta\\ -\dot\phi\sin\theta} \hspace{1cm} \text{and}\hspace{1cm} \dot {\bm{\omega}}^\mathcal{P}_{\mathcal{Q}\mathcal{P}} = \mymatrix{\ddot \phi \cos \theta - \dot\phi \dot \theta \sin \theta\\ - \ddot \theta \\ -\ddot \phi \sin \theta - \dot \phi\dot\theta\cos\theta}
\end{equation}

The equations of motion of the system are derived using a Newtonian approach. Under the assumption that the pendulum is subject only to its weight and the tension force ${\bm{N}}$, and exploiting the constant-length constraint $\ell$, which implies $\dot {\bm{r}}_{QP}^\mathcal{P} = \ddot {\bm{r}}_{QP}^\mathcal{P} = 0$, the absolute kinematics of the pendulum in the inertial frame are described by the following three vectorial equations
\begin{align}
{\bm{r}}_{OP}^\mathcal{I} & = {\bm{r}}_{OB}^\mathcal{I} + R_{\mathcal{I}\mathcal{B}}{\bm{r}}_{BQ}^\mathcal{B} + R_{\mathcal{I}\mathcal{P}}{\bm{r}}_{QP}^\mathcal{P}\\
\dot {\bm{r}}_{OP}^\mathcal{I} & = \dot {\bm{r}}_{OB}^\mathcal{I} + R_{\mathcal{I}\mathcal{B}}\crossmat{{\bm{\omega}}_{\mathcal{I}\mathcal{B}}^\mathcal{B}}{\bm{r}}_{BQ}^\mathcal{B} + R_{\mathcal{I}\mathcal{B}}\crossmat{{\bm{\omega}}_{\mathcal{I}\mathcal{B}}^\mathcal{B}}R_{\mathcal{B}\mathcal{P}}{\bm{r}}_{QP}^\mathcal{P} + R_{\mathcal{I}\mathcal{P}}\crossmat{{\bm{\omega}}_{\mathcal{Q}\mathcal{P}}^\mathcal{P}}{\bm{r}}_{QP}^\mathcal{P}\end{align}
\begin{equation}\begin{split}\ddot {\bm{r}}_{OP}^\mathcal{I} & = \ddot {\bm{r}}_{OB}^\mathcal{I} + R_{\mathcal{I}\mathcal{B}}\left(\crossmat{{\bm{\omega}}_{\mathcal{I}\mathcal{B}}^\mathcal{B}}^2 + \crossmat{\dot{\bm{\omega}}_{\mathcal{I}\mathcal{B}}^\mathcal{B}}\right){\bm{r}}_{BP}^\mathcal{B} + \\ & \hspace{3cm}+ 2R_{\mathcal{I}\mathcal{B}}\crossmat{{\bm{\omega}}_{\mathcal{I}\mathcal{B}}^\mathcal{B}}R_{\mathcal{B}\mathcal{P}}\crossmat{{\bm{\omega}}_{\mathcal{Q}\mathcal{P}}^\mathcal{P}}{\bm{r}}_{QP}^\mathcal{P} + R_{\mathcal{I}\mathcal{P}} \left(\crossmat{{\bm{\omega}}_{\mathcal{Q}\mathcal{P}}^\mathcal{P}}^2 + \crossmat{\dot {\bm{\omega}}_{\mathcal{Q}\mathcal{P}}^\mathcal{P}}\right){\bm{r}}_{QP}^\mathcal{P}\label{acckineq}
\end{split}\end{equation}
whereas the translational dynamics of the pendulum alone are given by
\begin{equation}m_P\ddot {\bm{r}}_{OP}^\mathcal{I} = m_P{\bm{g}}^\mathcal{I} + {\bm{N}}^\mathcal{I}\label{eqpnewt}\end{equation}
If we substitute the expression of $\ddot {\bm{r}}_{OP}^\mathcal{I}$ in Eq.~\eqref{acckineq} into Eq.~\eqref{eqpnewt}, and we project the resulting dynamics into $\mathcal{P}$ by pre-multiplying both sides by $R_{\mathcal{P}\mathcal{I}}$, Eq.~\eqref{eqpnewt} becomes
\begin{equation}\begin{split}
    & m_P R_{\mathcal{P}\mathcal{I}}\ddot {\bm{r}}_{OB}^\mathcal{I} - m_P R_{\mathcal{P}\mathcal{B}}\crossmat{{\bm{r}}_{BP}^\mathcal{B}}{\dot {\bm{\omega}}_{\mathcal{I}\mathcal{B}}^\mathcal{B}} - m_P \crossmat{{\bm{r}}_{QP}^\mathcal{P}} \dot {\bm{\omega}}_{\mathcal{Q}\mathcal{P}}^\mathcal{P} - {\bm{N}}^\mathcal{P} = \\ & \hspace{3cm} m_P {\bm{g}}^\mathcal{P} - m_P \left(2 \crossmat{{\bm{\omega}}_{\mathcal{I}\mathcal{B}}^\mathcal{P}} \crossmat{{\bm{\omega}}_{\mathcal{Q}\mathcal{P}}^\mathcal{P}} + \crossmat{{\bm{\omega}}_{\mathcal{Q}\mathcal{P}}^\mathcal{P}}^2\right){\bm{r}}_{QP}^\mathcal{P} - m_P R_{\mathcal{P}\mathcal{B}}\crossmat{{\bm{\omega}}_{\mathcal{I}\mathcal{B}}^\mathcal{B}}^2{\bm{r}}_{BP}^\mathcal{B}
\end{split}\label{pend3d}\end{equation}
We can now make the $\left(-m_P{\bm{r}}_{QP}^\mathcal{P} \times \dot {\bm{\omega}}_{\mathcal{Q}\mathcal{P}}^\mathcal{P} - {\bm{N}}^\mathcal{P}\right)$  term -- the only one that contains the pendulum's angular accelerations $\ddot \theta$ and $\ddot \phi$ and the tension $N$ -- explicit
\begin{equation}\begin{split}
    -m_P{\bm{r}}_{QP}^\mathcal{P} \times \dot {\bm{\omega}}_{\mathcal{Q}\mathcal{P}}^\mathcal{P} - {\bm{N}}^\mathcal{P} & = \mymatrix{m_P\ell \ddot \theta\\ m_P\ell\ddot \phi \cos \theta\\ -N} + \mymatrix{0\\ - m_P\ell\dot\phi \dot \theta \sin \theta\\ 0}
\end{split}\label{pendacc}\end{equation}
and rewrite the pendulum's dynamics as
\begin{equation}\begin{split}
    & m_P R_{\mathcal{P}\mathcal{I}}\ddot {\bm{r}}_{OB}^\mathcal{I} - m_P R_{\mathcal{P}\mathcal{B}}\crossmat{{\bm{r}}_{BP}^\mathcal{B}}{\dot {\bm{\omega}}_{\mathcal{I}\mathcal{B}}^\mathcal{B}} + m_P \ell \ddot \theta{\bm{e}}_x + m_P \ell\ddot \phi \cos \theta{\bm{e}}_y - N{\bm{e}}_z = m_P {\bm{g}}^\mathcal{P} + m_P{\bm{v}}
\end{split}\label{pend3d2}\end{equation}
where, by definition, 
\begin{equation}
    {\bm{v}} := \ell\dot \theta \dot \phi \sin \theta {\bm{e}}_y - R_{\mathcal{P}\mathcal{B}}\crossmat{{\bm{\omega}}_{\mathcal{I}\mathcal{B}}^\mathcal{B}}^2{\bm{r}}_{BP}^\mathcal{B} - \left(2 \crossmat{{\bm{\omega}}_{\mathcal{I}\mathcal{B}}^\mathcal{P}} \crossmat{{\bm{\omega}}_{\mathcal{Q}\mathcal{P}}^\mathcal{P}} + \crossmat{{\bm{\omega}}_{\mathcal{Q}\mathcal{P}}^\mathcal{P}}^2\right){\bm{r}}_{QP}^\mathcal{P}
\end{equation}
An important observation is that the tension force $N$ appears linearly and exclusively in the $z$-component of Eq.~\eqref{pend3d2}. Consequently, this equation can be solved explicitly for $N$, which can then be eliminated from the governing equations, reducing the number of variables required to describe the pendulum motion
\begin{equation}\begin{split}
N = m_P \left({\bm{z}}_\mathcal{P}^\mathcal{I}\right)^\top \ddot{\bm{r}}_{OB}^\mathcal{I} - m_P \left({\bm{z}}_\mathcal{P}^\mathcal{B}\right)^\top \crossmat{{\bm{r}}_{BP}^\mathcal{B}} \dot {\bm{\omega}}_{\mathcal{I}\mathcal{B}} ^\mathcal{B} - m_P g_z^\mathcal{P} - mv_z
\label{tensionpend}
\end{split}\end{equation}
Having derived the equations of motion of the pendulum, we now turn our attention to the dynamics of the rigid body to which it is attached. Assuming that both the rigid body and the pendulum are subjected to the same gravitational acceleration ${\bm{g}}$, the translational and rotational dynamics of the rigid-body center of mass are given by\footnote{For the torque expression, ${\bm{r}}_{BQ}$ could equivalently be used in place of ${\bm{r}}_{BP}$, since the tension force acts along the line joining $Q$ and $P$.}
\begin{align}
    m_B \ddot {\bm{r}}_{OB}^\mathcal{I} & = {\bm{F}}_B^\mathcal{I} + m_B{\bm{g}}^\mathcal{I} - N {\bm{z}}_\mathcal{P}^\mathcal{I}\label{rig3d1}\\ 
    J_B^\mathcal{B} \dot {\bm{\omega}}_{\mathcal{I}\mathcal{B}}^\mathcal{B} + {\bm{\omega}}_{\mathcal{I}\mathcal{B}}^\mathcal{B}\times J_B^\mathcal{B}{\bm{\omega}}_{\mathcal{I}\mathcal{B}}^\mathcal{B} & = {\bm{\tau}}^\mathcal{B} - {\bm{r}}_{BP}^\mathcal{B} \times N{\bm{z}}_\mathcal{P}^\mathcal{B}
\label{rig3d}\end{align}
Here, ${\bm{F}}_B$ denotes the external force acting at the rigid-body center of mass $B$, while ${\bm{\tau}}$ represents the externally applied torque. It is worth noting that the total torque acting about $B$ is given by the sum of the external torque and the moment generated by the pendulum's tension force, namely 
\[
{\bm{\tau}}_B = {\bm{\tau}} - {\bm{r}}_{BP} \times {\bm{N}}
\]
The $x$ and $y$ components of Eqs.~\eqref{pend3d2}, together with \eqref{rig3d1} and \eqref{rig3d}, fully describe the coupled pendulum--rigid body dynamics. 

By substituting Eq.~\eqref{tensionpend} into Eqs.~\eqref{pend3d2}, \eqref{rig3d1}, and \eqref{rig3d}, the tension force $N$ can be eliminated, yielding the final set of coupled nonlinear equations of motion. These can be written compactly in the form $\mathcal{M}\ddot {\bm{x}} = {\bm{u}}$ as\footnote{the last two equations in $\ddot \theta$ and $\ddot \phi$ can also be divided by the common term $m_P$. This is not done in this document to keep consistent units in the mass matrix, even though consistency is not strictly needed.}
\begin{equation}
\begin{split}
& \mymatrix{m_B\cdot I_{3\times 3} + m_P \left({\bm{z}}_\mathcal{P}^\mathcal{I} \otimes {\bm{z}}_\mathcal{P}^\mathcal{I}\right) & - m_P \left({\bm{z}}_\mathcal{P}^\mathcal{I} \otimes {\bm{z}}_\mathcal{P}^\mathcal{B}\right)\crossmat{{\bm{r}}_{BP}^\mathcal{B}} & {\bm{0}}_{3\times 1} & {\bm{0}}_{3\times 1} \\
m_P\crossmat{{\bm{r}}_{BP}^\mathcal{B}} \left({\bm{z}}_\mathcal{P}^\mathcal{B} \otimes {\bm{z}}_\mathcal{P}^\mathcal{I}\right) & J_B^\mathcal{B} - m_P \crossmat{{\bm{r}}_{BP}^\mathcal{B}} \left({\bm{z}}_\mathcal{P}^\mathcal{B} \otimes {\bm{z}}_{P}^\mathcal{B}\right)\crossmat{{\bm{r}}_{BP}^\mathcal{B}} & {\bm{0}}_{3\times 1} & {\bm{0}}_{3\times 1}\\ 
m_P \left({\bm{x}}_\mathcal{P}^\mathcal{I}\right)^\top & -m_P \left({\bm{x}}_\mathcal{P}^\mathcal{B}\right)^\top\crossmat{{\bm{r}}_{BP}^\mathcal{B}} & m_P \ell  & 0\\
m_P \left({\bm{y}}_\mathcal{P}^\mathcal{I}\right)^\top & -m_P\left({\bm{y}}_\mathcal{P}^\mathcal{B}\right)^\top\crossmat{{\bm{r}}_{BP}^\mathcal{B}}  & 0 & m_P \ell\cos \theta} 
\mymatrix{\ddot {\bm{r}}_{OB}^\mathcal{I}\\\dot {\bm{\omega}}_{\mathcal{I}\mathcal{B}}^\mathcal{B}\\\ddot \theta \\ \ddot \phi} \\ & \hspace{3.8cm} = \mymatrix{ {\bm{F}}_B^\mathcal{I} + m_B{\bm{g}}^\mathcal{I} + m_P\left(g_z^\mathcal{P} + v_z\right){\bm{z}}_\mathcal{P}^\mathcal{I}\\
{\bm{\tau}}^\mathcal{B} - {\bm{\omega}}_{\mathcal{I}\mathcal{B}}^\mathcal{B}\times J_B^\mathcal{B}{\bm{\omega}}_{\mathcal{I}\mathcal{B}}^\mathcal{B} + m_P\left(g_z^\mathcal{P} + v_z\right) {\bm{r}}_{BP}^\mathcal{B} \times {\bm{z}}_\mathcal{P}^\mathcal{B}\\
m_P\left(g_x^\mathcal{P} + v_x\right)\\m_P\left(g_y^\mathcal{P} + v_y\right)}
\end{split}
\label{multibred}\end{equation}
The global mass matrix $\mathcal{M}$ and the input vector ${\bm{u}}$ depend on  the external forces and torques, the gravitational acceleration, and the state of the system. For numerical simulation purposes, Eq.~\eqref{multibred} can be solved for the acceleration vector $\ddot {\bm{x}}$, which can subsequently be integrated together with the kinematics to obtain the temporal evolution of the state. 

It is important to note that $\mathcal{M}$ remains nonsingular only when $\cos \theta \ne 0$. Consequently, the present formulation becomes singular for $\left|\theta\right|$, implying that the pendulum angle must remain within this limit for the model to be numerically well posed. Although this represents a mathematical limitation of the chosen coordinates, it should not be regarded as a practical restriction. Indeed, sloshing amplitudes approaching such values would indicate a severe degradation of the system behaviour rather than a deficiency of the model itself. In realistic spacecraft applications, sloshing angles are typically expected to remain within a few tens of degrees at most, which also corresponds to the range over which equivalent mechanical models provide a meaningful approximation of the more complex fluid dynamics of liquid propellants.

\paragraph{Adding damping}
A simple viscous damping model can be incorporated by introducing a damping coefficient $q$, expressed in [kg m/s], acting on the angular rates of the pendulum. The pendulum equations of motion (e.g., Eq.~\eqref{multibred}) then become
\begin{align}
    m_P\ell \ddot \theta + q \dot \theta & = \dots\\
    m_P\ell \ddot \phi \cos\theta + q \dot \phi & = \dots
\end{align}
The damping forces associated with these terms act in directions orthogonal to ${\bm{z}}_\mathcal{P}$, and therefore do not contribute to the tension force $N$. As a consequence, the expression for $N$ remains unchanged, and no additional terms appear in the rigid-body equations of motion. The damping affects the coupled dynamics only through the pendulum equations, while all other equations remain unaltered.

\subsection{Multiple pendulums}\label{secmultipen}
Inspection of the expressions of $\mathcal{M}$ and ${\bm{u}}$ in Eq.~\eqref{multibred} reveals a clear separation between the terms associated with the rigid-body dynamics of the spacecraft and those describing the pendulum motion. This observation naturally extends the formulation to systems comprising multiple pendulums, representing, for example, multiple propellant tanks and/or multiple sloshing modes within a single tank.
For a system with $n$ pendulums, the generalized acceleration vector can be defined as $\ddot {\bm{x}} = \left(\ddot {\bm{r}}_{OB}^\mathcal{I}, \dot {\bm{\omega}}_{\mathcal{I}\mathcal{B}}^\mathcal{B}, \ddot \theta_1, \ddot \phi_1,\dots, \ddot \theta_n, \ddot \phi_n\right)$, where each pair $\left(\theta_i, \phi_i\right)$ describes the motion of the $i$-th pendulum. The corresponding mass matrix $\mathcal{M}$ and input vector ${\bm{u}}$ assume the following block structure:
\begin{equation}
    \mathcal{M} = \left[\begin{array}{c|ccc}
    M_R  + \sum_i A_i& {0_{6\times 2}} & \dots & {0_{6\times 2}}\\ \hline 
    B_1 & C_1 & \\
    \vdots & &\ddots\\
    B_n &  &  & C_n\end{array}\right]
\hspace{2cm}
{\bm{u}} = \mymatrix{{\bm{u}}_R + \sum_i{\bm{a}}_i\\ \hline {\bm{b}}_1\\ \vdots\\{\bm{b}}_n}
\label{massstructmulti}
\end{equation}
where $M_R$ is the rigid mass matrix of the spacecraft (i.e., without pendulums' masses) and ${\bm{u}}_R$ the input torsor, defined as
\begin{equation}
    M_R := \mymatrix{m_B \cdot I_{3\times 3} & 0_{3\times 3} \\ 0_{3\times 3} & J_B^\mathcal{B}}\hspace{2cm}
    {\bm{u}}_R := \mymatrix{{\bm{F}}_B^\mathcal{I} + m_B{\bm{g}}^\mathcal{I} \\{\bm{\tau}}^\mathcal{B} - {\bm{\omega}}_{\mathcal{I}\mathcal{B}}^\mathcal{B}\times J_B^\mathcal{B}{\bm{\omega}}_{\mathcal{I}\mathcal{B}}^\mathcal{B}}
    \label{urdef}
\end{equation}
The auxiliary matrices $A_i$, $B_i$, $C_i$, and vectors ${\bm{a}}_i$, ${\bm{b}}_i$, ${\bm{v}}_i$ depend on the properties of the $i$-th pendulum, and are defined as follows
\begin{align}
A_i &= m_i\mymatrix{{\bm{z}}_{\mathcal{P}_i}^\mathcal{I} \otimes {\bm{z}}_{\mathcal{P}_i}^\mathcal{I} & - \left({\bm{z}}_{\mathcal{P}_i}^\mathcal{I} \otimes {\bm{z}}_{\mathcal{P}_i}^\mathcal{B}\right)\crossmat{{\bm{r}}_{BP_i}^\mathcal{B}}\label{defpendAi}\\
\crossmat{{\bm{r}}_{BP_i}^\mathcal{B}} \left({\bm{z}}_{\mathcal{P}_i}^\mathcal{B} \otimes {\bm{z}}_{\mathcal{P}_i}^\mathcal{I}\right) & - \crossmat{{\bm{r}}_{BP_i}^\mathcal{B}} \left({\bm{z}}_{\mathcal{P}_i}^\mathcal{B} \otimes {\bm{z}}_{\mathcal{P}_i}^\mathcal{B}\right)\crossmat{{\bm{r}}_{BP_i}^\mathcal{B}}} \\
B_i &  = m_i\mymatrix{\left({\bm{x}}_{\mathcal{P}_i}^\mathcal{I}\right)^\top & -\left({\bm{x}}_{\mathcal{P}_i}^\mathcal{B}\right)^\top\crossmat{{\bm{r}}_{B{P_i}}^\mathcal{B}} \\
\left({\bm{y}}_{\mathcal{P}_i}^\mathcal{I}\right)^\top & -\left({\bm{y}}_{\mathcal{P}_i}^\mathcal{B}\right)^\top\crossmat{{\bm{r}}_{B{P_i}}^\mathcal{B}}}\label{defpendBi} \\
C_i &:= m_i\ell_i\mymatrix{1 & 0\\ 0 & \cos\theta_i}\label{defpendCi}\\
{\bm{a}}_i & := m_i\left(g_z^{\mathcal{P}_i} + v_{i,z}\right)\mymatrix{{\bm{z}}_{\mathcal{P}_i}^\mathcal{I} \label{defpendai}\\
{\bm{r}}_{BP_i}^{B}\times{\bm{z}}_{\mathcal{P}_i}^\mathcal{B}}\\
{\bm{b}}_i & := m_i \mymatrix{g_x^{\mathcal{P}_i} + v_{i, x}\\g_y^{\mathcal{P}_i} + v_{i, y}}\\
{\bm{v}}_i & := \ell_i\dot \theta_i \dot \phi_i \sin \theta_i {\bm{e}}_y - R_{P_iB}\crossmat{{\bm{\omega}}_{\mathcal{I}\mathcal{B}}^\mathcal{B}}^2{\bm{r}}_{BP_i}^\mathcal{B} - \left(2 \crossmat{{\bm{\omega}}_{\mathcal{I}\mathcal{B}}^{\mathcal{P}_i}} \crossmat{{\bm{\omega}}_{Q_iP_i}^{\mathcal{P}_i}} + \crossmat{{\bm{\omega}}_{Q_iP_i}^{\mathcal{P}_i}}^2\right){\bm{r}}_{QP_i}^{\mathcal{P}_i}
\end{align}
The resulting linear system $\mathcal{M}\ddot {\bm{x}} = {\bm{u}}$ can be efficiently solved for the accelerations vector by exploiting the sparsity of $\mathcal{M}$. Since all of the $C_i$'s are diagonal matrices, we only need to solve for a $6 \times 6$ linear system to first retrieve $\ddot {\bm{r}}_{OB}^\mathcal{I}$ and $\dot {\bm{\omega}}_{\mathcal{I}\mathcal{B}}^\mathcal{B}$
\begin{equation}
    \mymatrix{\ddot {\bm{r}}_{OB}^\mathcal{I}\\\dot {\bm{\omega}}_{\mathcal{I}\mathcal{B}}^\mathcal{B}} = \left(M_R + \sum_i A_i\right)^{-1}\left({\bm{u}}_R + \sum_i {\bm{a_i}}\right)
\end{equation}
and then compute the pendulums' angular accelerations as
\begin{equation}
    \mymatrix{\ddot \theta_i \\ \ddot \phi_i} = \frac{1}{m_i\ell_i}\mymatrix{1 & 0\\ 0 & 1/\cos\theta_i}\left({\bm{b}}_i - B_i\mymatrix{\ddot {\bm{r}}_{OB}^\mathcal{I}\\\dot {\bm{\omega}}_{\mathcal{I}\mathcal{B}}^\mathcal{B}}\right)
\end{equation}

\paragraph{Pendulum forces and torques}
The force and torque generated by the $i$-th pendulum and applied to the rigid body center of mass $B$ are explicitly shown in Eqs.~\eqref{rig3d1} and \eqref{rig3d} (for a single pendulum), and are given by
\[{\bm{u}}_{i} := \mymatrix{{\bm{F}}_{Bi}^\mathcal{I}\\{\bm{\tau}}_{Bi}^\mathcal{B}\\} = -N_i\mymatrix{{\bm{z}}_{\mathcal{P}_i}^\mathcal{I}\\ {\bm{r}}_{BP_i}^\mathcal{B} \times {\bm{z}}_{\mathcal{P}_i}^\mathcal{B}}\]
Using the definitions introduced in Eqs.~\eqref{defpendAi} and \eqref{defpendai}, ${\bm{u}}_{i}$ can be also directly computed as
\begin{equation}
    {\bm{u}}_{i} = {\bm{a}}_i - A_i\mymatrix{\ddot {\bm{r}}_{OB}^\mathcal{I}\\ \dot {\bm{\omega}}_{\mathcal{I}\mathcal{B}}^\mathcal{B}}
\end{equation}

\newpage
\section{Linearized pendulum model}
In the previous section, the nonlinear equations of motion were derived using the absolute position of the rigid body's center of mass, ${\bm{r}}^\mathcal{I}_{OB}$, as a state variable. While this formulation is convenient for deriving the system dynamics, it is not necessarily the most suitable choice for controller design, analysis, and implementation. In practice, the onboard software is expected to estimate the system center of mass under the assumption of a rigidly distributed propellant mass, thereby neglecting sloshing effects, and to regulate its position and velocity through the application of control forces and torques. Consequently, when deriving a linearized model, it is preferable to formulate the dynamics with respect to a nominal global center of mass rather than the rigid-body center of mass $B$. This nominal point, denoted by $\bar{G}$, corresponds to the overall center of mass of the rigid body and pendulum when $\theta=\phi=0$. With this definition, $\bar{G}$ remains fixed in the body reference frame, providing a more natural representation for guidance and control purposes.

\subsection{Nonlinear model at the nominal CoM}
We now derive the nonlinear equations of motion using the nominal global center of mass $\bar G$ -- defined as the center of mass of the system for $\theta=\phi=0$ -- to describe the translational dynamics. This formulation exploits the fact that both ${\bm{r}}{B\bar G}^\mathcal{B}$ and ${\bm{r}}{\bar GQ}^\mathcal{B}$ are constant in the body-fixed frame. As in the previous derivation, the dynamics of the rigid body and the pendulum are treated separately. However, linear and angular momentum balances are now formulated with respect to $\bar G$, which serves as the common reference point, yielding
\begin{align}
m_B \ddot {\bm{r}}_{O\bar G}^\mathcal{I} + m_B R_{\mathcal{I}\mathcal{B}}\crossmat{{\bm{r}}_{B\bar G}^\mathcal{B}} \dot {\bm{\omega}}_{\mathcal{I}\mathcal{B}}^\mathcal{B} - m_B R_{\mathcal{I}\mathcal{B}}\crossmat{{\bm{\omega}}_{\mathcal{I}\mathcal{B}}^\mathcal{B}}^2 {\bm{r}}_{B\bar G}^\mathcal{B} = {\bm{F}}_{\bar G}^\mathcal{I} + m_B {\bm{g}}^\mathcal{I} - N {\bm{z}}_\mathcal{P}^\mathcal{I}\\
-m_B \crossmat{{\bm{r}}_{B\bar G}^\mathcal{B}} R_{\mathcal{B}\mathcal{I}}\ddot {\bm{r}}_{O\bar G}^\mathcal{I} + J_{\bar G}^\mathcal{B} \dot {\bm{\omega}}_{\mathcal{I}\mathcal{B}}^\mathcal{B} + {\bm{\omega}}_{\mathcal{I}\mathcal{B}}^\mathcal{B} \times J_{\bar G}^\mathcal{B} {\bm{\omega}}_{\mathcal{I}\mathcal{B}}^\mathcal{B} = {\bm{\tau}}^\mathcal{B} - {\bm{r}}_{B \bar G}^\mathcal{B} \times m_B {\bm{g}}^\mathcal{B} - {\bm{r}}_{\bar G P}^\mathcal{B} \times N {\bm{z}}_{P}^\mathcal{B}
\end{align}
Here, $J_{\bar G}^\mathcal{B}$ denotes the inertia tensor of the rigid body about $\bar G$. It is worth noting that, since the rigid-body weight is modelled as acting at $B$, it generates a gravitational torque about $\bar G$. The pendulum dynamics remain unchanged with respect to Eq.~\eqref{pend3d2}; the only modification is that every occurrence of ${\bm{r}}_{BP}^{\mathcal{B}}$ must be replaced by ${\bm{r}}_{\bar G P}^{\mathcal{B}}$. The complete nonlinear equations of motion become
\small
\begin{equation}
\begin{split}
& \mymatrix{m_B\cdot I_{3\times 3} + m_P \left({\bm{z}}_\mathcal{P}^\mathcal{I} \otimes {\bm{z}}_\mathcal{P}^\mathcal{I}\right) & \textcolor{red}{m_B R_{\mathcal{I}\mathcal{B}}\crossmat{{\bm{r}}_{B\bar G}^\mathcal{B}}} - m_P \left({\bm{z}}_\mathcal{P}^\mathcal{I} \otimes {\bm{z}}_\mathcal{P}^\mathcal{B}\right)\crossmat{{\bm{r}}_{\bar G P}^\mathcal{B}} & {\bm{0}}_{3\times 1} & {\bm{0}}_{3\times 1} \\
\textcolor{red}{-m_B \crossmat{{\bm{r}}_{B\bar G}^\mathcal{B}} R_{\mathcal{B}\mathcal{I}}} + m_P\crossmat{{\bm{r}}_{\bar G P}^\mathcal{B}} \left({\bm{z}}_\mathcal{P}^\mathcal{B} \otimes {\bm{z}}_\mathcal{P}^\mathcal{I}\right) & J_{\bar G}^\mathcal{B} - m_P \crossmat{{\bm{r}}_{\bar G P}^\mathcal{B}} \left({\bm{z}}_\mathcal{P}^\mathcal{B} \otimes {\bm{z}}_{P}^\mathcal{B}\right)\crossmat{{\bm{r}}_{\bar G P}^\mathcal{B}} & {\bm{0}}_{3\times 1} & {\bm{0}}_{3\times 1}\\ 
m_P \left({\bm{x}}_\mathcal{P}^\mathcal{I}\right)^\top & -m_P \left({\bm{x}}_\mathcal{P}^\mathcal{B}\right)^\top\crossmat{{\bm{r}}_{\bar G P}^\mathcal{B}} & m_P \ell  & 0\\
m_P \left({\bm{y}}_\mathcal{P}^\mathcal{I}\right)^\top & -m_P\left({\bm{y}}_\mathcal{P}^\mathcal{B}\right)^\top\crossmat{{\bm{r}}_{\bar G P}^\mathcal{B}}  & 0 & m_P \ell\cos \theta} 
\mymatrix{\ddot {\bm{r}}_{O\bar G}^\mathcal{I}\\\dot {\bm{\omega}}_{\mathcal{I}\mathcal{B}}^\mathcal{B}\\\ddot \theta \\ \ddot \phi} \\ & \hspace{3.8cm} = \mymatrix{ \textcolor{red}{m_B R_{\mathcal{I}\mathcal{B}}\crossmat{{\bm{\omega}}_{\mathcal{I}\mathcal{B}}^\mathcal{B}}^2 {\bm{r}}_{B\bar G}^\mathcal{B}} + {\bm{F}}_{\bar G}^\mathcal{I} + m_B{\bm{g}}^\mathcal{I} + m_P\left(g_z^\mathcal{P} + v_z\right){\bm{z}}_\mathcal{P}^\mathcal{I}\\
{\bm{\tau}}^\mathcal{B} \textcolor{red}{- m_B {\bm{r}}_{B \bar G}^\mathcal{B} \times {\bm{g}}^\mathcal{B}} - {\bm{\omega}}_{\mathcal{I}\mathcal{B}}^\mathcal{B}\times J_{\bar G}^\mathcal{B}{\bm{\omega}}_{\mathcal{I}\mathcal{B}}^\mathcal{B} + m_P\left(g_z^\mathcal{P} + v_z\right) {\bm{r}}_{\bar G P}^\mathcal{B} \times {\bm{z}}_\mathcal{P}^\mathcal{B}\\
m_P\left(g_x^\mathcal{P} + v_x\right)\\m_P\left(g_y^\mathcal{P} + v_y\right)}
\end{split}\label{nonlininG}
\end{equation}
\normalsize
Comparing the above formulation with Eq.~\eqref{multibred}, a number of additional terms can be identified in the equations of motion; these contributions are highlighted in \textcolor{red}{red}. All remaining terms retain the same mathematical structure as those in Eq.~\eqref{multibred}. It should be noted, however, that the rigid-body inertia, state variables, external forces, and pendulum position are now defined with respect to $\bar G$. Consequently, although the equations are formally identical, their physical interpretation differs from that of Eq.~\eqref{multibred}. Before proceeding with the linearization, it is convenient to exploit the definition of the nominal center of mass $\bar G$, wich implies that
\[\begin{split}
    m_B{\bm{r}}_{\bar G B} + m_P\bar {\bm{r}}_{\bar G P} = 0
\end{split}\]
From this relation, it follows that
\begin{equation}
    {\bm{r}}_{B\bar G} = \frac{m_P}{m_B}\bar {\bm{r}}_{\bar G P}\label{comrela}
\end{equation}
which will prove useful in simplifying several terms of the linearized equations of motion.

\subsection{Single pendulum: force in inertial frame} 
The coupled equations of motion in Eq.~\eqref{nonlininG} can be readily linearized\footnote{throughout the linearization process, $\bar p$ denotes the nominal value of a quantity $p$, while $\delta p$ denotes its infinitesimal perturbation, such that $p = \bar p + \delta p$.} about an equilibrium characterized by a constant spacecraft attitude aligned with the inertial frame, i.e., $\bar R_{\mathcal{I}\mathcal{B}} = I_{3\times 3}$, and a prescribed nominal translational trajectory driven by gravity and a constant longitudinal force $\bar {\bm{F}}_{\bar G}^\mathcal{I} = \bar F_z {\bm{e}}_z$ with $\bar F_z > 0$. Without loss of generality, the body-fixed frame is chosen to be aligned with the fulcrum frame, yielding $R_{BQ} = I_{3\times 3}$ and simplifying the resulting equations. Under these assumptions, the nominal pendulum configuration corresponds to $\bar \theta = \bar \phi = 0$. Consequently, $\bar R_{\mathcal{B}\mathcal{P}} = I_{3\times 3}$, $\bar {\bm{z}}_\mathcal{P}^\mathcal{I} = \bar {\bm{z}}_\mathcal{P}^\mathcal{B} = {\bm{e}}_z$ and similarly for the remaining axes. Small perturbations of the body and pendulum attitudes can therefore be described using a first-order small-angle approximation, such that
\begin{align}
    \delta R_{\mathcal{I}\mathcal{B}} & = \crossmat{\delta{\bm{\theta}}_{\mathcal{I}\mathcal{B}}} = \mymatrix{0 & -\delta \theta_{\mathcal{I}\mathcal{B}z} & \delta \theta_{\mathcal{I}\mathcal{B}y}\\ \delta \theta_{\mathcal{I}\mathcal{B}z} & 0 & -\delta \theta_{\mathcal{I}\mathcal{B}x}\\ -\delta \theta_{\mathcal{I}\mathcal{B}y} & \delta \theta_{\mathcal{I}\mathcal{B}x} & 0}\\
    \delta R_{\mathcal{B}\mathcal{P}} & = \mymatrix{0 & 0 & -\delta \theta\\0 & 0 & -\delta \phi\\ \delta \theta & \delta \phi & 0}\\
    \delta R_{\mathcal{P}\mathcal{I}} & = - \delta R_{\mathcal{I}\mathcal{B}} - \delta R_{\mathcal{B}\mathcal{P}}
\end{align}
and so
\[\begin{split}
\delta {\bm{z}}_\mathcal{P}^\mathcal{B} = \delta R_{\mathcal{B}\mathcal{P}}{\bm{e}}_z &\hspace{1cm}\text{and}\hspace{1cm}\delta {\bm{z}}_\mathcal{P}^\mathcal{I} = -\delta R_{\mathcal{P}\mathcal{I}}{\bm{e}}_z\\
\delta {\bm{g}}^\mathcal{B} = \delta R_{\mathcal{B}\mathcal{I}}{\bm{g}}^\mathcal{I}\\
\delta g_x^\mathcal{P} = {\bm{e}}_x^\top \delta R_{\mathcal{P}\mathcal{I}}{\bm{g}}^\mathcal{I}&\hspace{1cm}\text{and similarly for $g_y^\mathcal{P}$ and $g_z^\mathcal{P}$}\\
\delta {\bm{r}}_{\bar GP}^\mathcal{B} = -\ell \delta {\bm{z}}_\mathcal{P}^\mathcal{B}&
\end{split}\]
The perturbed dynamics can be expressed as
\[\left(\bar{\mathcal{M}} + \delta \mathcal{M}\right)\left(\ddot{\bar{{\bm{x}}}} + \delta \ddot {\bm{x}}\right) = \bar{\bm{u}} + \delta {\bm{u}}\]
from which the nominal dynamics and the first-order perturbation dynamics are obtained as
\begin{align}
    \bar{\mathcal{M}}\ddot{\bar{{\bm{x}}}} & = \bar{\bm{u}}\\
    \bar{\mathcal{M}}\delta \ddot {\bm{x}} + \delta \mathcal{M}\ddot{\bar{{\bm{x}}}} & \approx \delta {\bm{u}}\label{coupllindin}
\end{align}
We assume that a nominal control torque $\bar{\bm{\tau}}^\mathcal{B}$ is applied to the rigid body to exactly compensate the torque generated by the pendulum, thereby ensuring zero nominal angular velocity and angular acceleration. Under the adopted linearization assumptions, namely a purely longitudinal nominal force and a constant attitude aligned with the inertial frame, the nominal acceleration vector becomes
\begin{equation}
    \ddot {\bar{{\bm{x}}}} = \mymatrix{\ddot {\bar{\bm{r}}}_{O\bar G}^\mathcal{I}\\ {\bm{0}}_{5\times 1}} = \mymatrix{{\bm{g}}^\mathcal{I} + \frac{\bar F_z}{m_B + m_P} {\bm{e}}_z\\{\bm{0}}_{5\times 1}}
\end{equation}
It is worth noting that $\ddot {\bar{\bm{r}}}_{O\bar G}^\mathcal{I}$ is generally non-zero, as it is driven by both the gravitational acceleration (which can have any direction in $I$) and the nominal force applied to the system. Only the rotational dynamics and the pendulum motion vanish at the equilibrium point.

Exploiting the equilibrium conditions $\bar{\bm{v}} = \delta {\bm{v}} = 0$ and the relation $\bar {\bm{r}}_{\bar GP}^\mathcal{B} = {\bm{r}}_{\bar GQ}^\mathcal{B} - \ell {\bm{e}}_z$, the various terms appearing in the linearized equations of motion can be readily evaluated. In particular
\footnotesize
\begin{align}
\delta \mathcal{M} & = \mymatrix{m_P \left(\delta {\bm{z}}_\mathcal{P}^\mathcal{I} \otimes \bar{\bm{z}}_\mathcal{P}^\mathcal{I} + \bar {\bm{z}}_\mathcal{P}^\mathcal{I} \otimes \delta {\bm{z}}_\mathcal{P}^\mathcal{I}\right) & \dots  \\
m_B \left(\crossmat{{\bm{r}}_{B\bar G}^\mathcal{B}}\delta R_{\mathcal{I}\mathcal{B}} - \crossmat{\delta {\bm{r}}_{B \bar G}^\mathcal{B}} \right) + m_P\left\{\crossmat{\delta {\bm{r}}_{\bar GP}^\mathcal{B}} \left(\bar {\bm{z}}_\mathcal{P}^\mathcal{B} \otimes \bar {\bm{z}}_\mathcal{P}^\mathcal{I}\right) + \crossmat{\bar {\bm{r}}_{\bar GP}^\mathcal{B}} \left(\delta {\bm{z}}_\mathcal{P}^\mathcal{B} \otimes \bar {\bm{z}}_\mathcal{P}^\mathcal{I} + \bar {\bm{z}}_\mathcal{P}^\mathcal{B} \otimes \delta {\bm{z}}_\mathcal{P}^\mathcal{I}\right)\right\} &\dots \\ 
m_P\left(\delta {\bm{x}}_\mathcal{P}^\mathcal{I}\right)^\top & \dots\\
m_P\left(\delta {\bm{y}}_\mathcal{P}^\mathcal{I}\right)^\top & \dots}\\
\delta {\bm{u}} & = \mymatrix{\delta {\bm{F}}_{\bar G}^\mathcal{I} + m_P\left(\delta g_z^\mathcal{P} \bar {\bm{z}}_\mathcal{P}^\mathcal{I} + \bar g_z^\mathcal{P} \delta {\bm{z}}_\mathcal{P}^\mathcal{I}\right)\\
\delta {\bm{\tau}}^\mathcal{B} - m_B \left(\delta {\bm{r}}_{B\bar G}^\mathcal{B} \times {\bm{g}}^\mathcal{I} + {\bm{r}}_{B\bar G}^\mathcal{B} \times \delta {\bm{g}}^\mathcal{B}\right) + m_P\left(\delta g_z^\mathcal{P} \bar {\bm{r}}_{\bar GP}^\mathcal{B}\times \bar {\bm{z}}_\mathcal{P}^\mathcal{B} + \bar g_z^\mathcal{P} \delta {\bm{r}}_{BP}^\mathcal{B}\times \bar {\bm{z}}_\mathcal{P}^\mathcal{B} + \bar g_z^\mathcal{P} \bar {\bm{r}}_{\bar GP}^\mathcal{B}\times \delta {\bm{z}}_\mathcal{P}^\mathcal{B}\right)\\ m_P \delta g_x^\mathcal{P} \\ m_P \delta g_y^\mathcal{P}}
\end{align}
\normalsize
so that the coupled linearized dynamics in Eq.~\eqref{coupllindin} can be rewritten in the form of a second-order linear MIMO system $M\delta \ddot {\bm{x}} + K\delta {\bm{x}} = \delta {\bm{u}}_{ext}$
\small 
\begin{equation}\begin{split}\mymatrix{
m_B & 0 & 0  & 0 & -m_P \bar z_{\bar GP}^\mathcal{B} & m_P y_{\bar GQ}^\mathcal{B} & 0 & 0\\
0 & m_B & 0 & m_P \bar z_{\bar GP}^\mathcal{B} & 0 &-m_P x_{\bar GQ}^\mathcal{B} & 0 & 0\\
0& 0 & m_B+m_P & 0 & 0 & 0 & 0 & 0\\
0 & m_P \bar z_{\bar GP}^\mathcal{B} & 0 & J_{xx} + m_P (y_{\bar GQ}^\mathcal{B})^2 & J_{xy} - m_P x_{\bar GQ}^\mathcal{B} y_{\bar GQ}^\mathcal{B} & J_{xz} & 0 & 0\\
-m_P \bar z_{\bar GP}^\mathcal{B} & 0 & 0 & J_{xy} - m_P x_{\bar GQ}^\mathcal{B} y_{\bar GQ}^\mathcal{B} & J_{yy} + m_P (x_{\bar GQ}^\mathcal{B})^2 & J_{yz} & 0 & 0\\
m_P y_{\bar GQ}^\mathcal{B} & -m_P x_{\bar GQ}^\mathcal{B} & 0 & J_{xz} & J_{yz} & J_{zz} & 0 & 0\\
m_P & 0 & 0 & 0 & m_P\bar z_{\bar GP}^\mathcal{B} & -m_P y_{\bar GQ}^\mathcal{B} & m_P\ell & 0\\
0 & m_P & 0 & -m_P\bar z_{\bar GP}^\mathcal{B} & 0 & m_P x_{\bar GQ}^\mathcal{B} & 0 & m_P\ell}\mymatrix{\delta \ddot x_{O\bar G}^\mathcal{I}\\ \delta \ddot y_{O\bar G}^\mathcal{I}\\ \delta \ddot z_{O\bar G}^\mathcal{I} \\ \delta \ddot \theta_{\mathcal{I}\mathcal{B}x}\\ \delta \ddot \theta_{\mathcal{I}\mathcal{B}y}\\ \delta \ddot \theta_{\mathcal{I}\mathcal{B}z}\\ \delta \ddot \theta \\ \delta \ddot \phi} + \\
 + \frac{\bar F_z}{m_B + m_P}\mymatrix{
    0 & 0 & 0 & 0 & m_P & 0 & -m_P & 0\\
 0 & 0 & 0 & -m_P & 0 & 0 & 0 & -m_P\\
 0 & 0 & 0 & 0 & 0 & 0 & 0 & 0\\
 0 & 0 & 0 & m_P \bar z_{\bar GP}^\mathcal{B} & 0 & 0 & 0 & m_P z_{\bar GQ}^\mathcal{B}\\
 0 & 0 & 0 & 0 & m_P \bar z_{\bar GP}^\mathcal{B} & 0 & -m_P z_{\bar GQ}^\mathcal{B} & 0\\
 0 & 0 & 0 & -m_P x_{\bar GQ}^\mathcal{B} & -m_P y_{\bar GQ}^\mathcal{B} & 0 & m_P y_{\bar GQ}^\mathcal{B} & -m_P x_{\bar GQ}^\mathcal{B}\\
 0 & 0 & 0 & 0 & -m_P & 0 & m_P & 0\\
 0 & 0 & 0 & m_P & 0 & 0 & 0 & m_P}\mymatrix{\delta  x_{O\bar G}^\mathcal{I}\\ \delta  y_{O\bar G}^\mathcal{I}\\ \delta  z_{O\bar G}^\mathcal{I}\\ \delta \theta_{\mathcal{I}\mathcal{B}x}\\ \delta  \theta_{\mathcal{I}\mathcal{B}y}\\ \delta  \theta_{\mathcal{I}\mathcal{B}z}\\ \delta  \theta \\ \delta  \phi} = \mymatrix{\delta F_{\bar Gx}^\mathcal{I}\\\delta F_{\bar Gy}^\mathcal{I}\\\delta F_{\bar Gz}^\mathcal{I}\\\delta \tau_x^\mathcal{B}\\\delta \tau_y^\mathcal{B}\\\delta \tau_z^\mathcal{B}\\0\\0}
\end{split}\label{linmodcomplete}\end{equation}
\normalsize
Thanks to the particular choice of pendulum angular coordinates, the dynamics of $\delta \theta$ and $\delta \phi$ are decoupled. As a result, the linearized three-dimensional pendulum dynamics can be interpreted as the superposition of two independent planar pendulums. It is convenient to note that the nominal force and torque applied to the rigid body are
\begin{align}
    \bar F_z & = \left(m_B + m_P\right)\left(\ddot{\bar{z}}_{O\bar G}^\mathcal{I} - g_z^\mathcal{I}\right)\\
    \bar {\bm{\tau}}^\mathcal{B} & = 0
\end{align}

\paragraph{Modal coordinates}
Starting from the eight scalar equations of motion in Eq.~\eqref{linmodcomplete}, we can define a new set of degrees of freedom $\eta_\theta$ and $\eta_\phi$, such that
\begin{align}
    \eta_\theta & := \sqrt{m_P}\ell \left(\delta \theta - \delta \theta_{\mathcal{I}\mathcal{B}y}\right) \\
    \eta_\phi & := \sqrt{m_P}\ell \left(\delta \phi + \delta \theta_{\mathcal{I}\mathcal{B}x}\right)
\end{align}
and sequentially perform the following algebraic operations on the equations of motion:
\begin{enumerate}
\item Apply inverse coordinate transformation, to transition from $(\delta \theta, \delta \phi)$ to $(\eta_\theta, \eta_\phi)$.
 \item Replace Eq.~1 by the sum of Eq.~1 and Eq.~7.
 \item Replace Eq.~2 by the sum of Eq.~2 and Eq.~8.
 \item Multiply Eq.~8 by $-z_{\bar G Q}^\mathcal{B}$ and add it to Eq.~4.
 \item Multiply Eq.~7 by $+z_{\bar G Q}^\mathcal{B}$ and add it to Eq.~5.
 \item Multiply Eq.~7 by $-y_{\bar G Q}^\mathcal{B}$, multiply Eq.~8 by $x_{\bar G Q}^\mathcal{B}$ and add both of them to Eq.~6.
 \item Divide Eq.~7 and Eq.~8 by $\sqrt{m_P}$.
\end{enumerate}
The resulting dynamics are now written in the so-called \emph{modal} form
\small
\begin{equation}\begin{split}
\left[\begin{array}{cccccc|cc}
m_{\bar G}  & 0 & 0 & 0 & m_P \ell & 0 & \sqrt{m_P}  & 0\\
0 & m_{\bar G} & 0 &  - m_P \ell & 0 & 0 & 0 & \sqrt{m_P}\\
0 & 0 &  m_{\bar G}& 0 & 0  & 0 &  0 & 0\\
0 & - m_P \ell &  0 & I_{xx}^\mathcal{B}  & I_{xy}^\mathcal{B} &  I_{xz}^\mathcal{B} & 0 & -\sqrt{m_P}z_{\bar GQ}^\mathcal{B}\\
m_P \ell & 0 & 0 & I_{xy}^\mathcal{B}  & I_{yy}^\mathcal{B} &  I_{yz}^\mathcal{B} & \sqrt{m_P} z_{\bar GQ}^\mathcal{B} & 0\\
0 & 0  & 0 & I_{xz}^\mathcal{B} & I_{yz}^\mathcal{B} &  I_{zz}^\mathcal{B}  & -\sqrt{m_P} y_{\bar GQ}^\mathcal{B} &  \sqrt{m_P} x_{\bar GQ}^\mathcal{B}\\ \hline
\sqrt{m_P} & 0 & 0 & 0 & \sqrt{m_P}z_{\bar GQ}^\mathcal{B}  & -\sqrt{m_P} y_{\bar GQ}^\mathcal{B}  & 1 & 0\\
0 & \sqrt{m_P} & 0 & -\sqrt{m_P} z_{\bar GQ}^\mathcal{B} & 0 & \sqrt{m_P} x_{\bar GQ}^\mathcal{B} & 0 & 1
\end{array}\right]\mymatrix{\delta  \ddot x_{O\bar G}^\mathcal{I}\\ \delta  \ddot y_{O\bar G}^\mathcal{I}\\ \delta \ddot  z_{O\bar G}^\mathcal{I}\\ \delta \ddot \theta_{\mathcal{I}\mathcal{B}x}\\ \delta \ddot  \theta_{\mathcal{I}\mathcal{B}y}\\ \delta \ddot  \theta_{\mathcal{I}\mathcal{B}z}\\ \hline \ddot \eta_\theta \\ \ddot \eta_\phi} + \\
+ 
\left[\begin{array}{cccccc|cc}
0 & 0 & 0 & 0 & 0 & 0 & 0 & 0\\
0 & 0 & 0 & 0 & 0 & 0 &  0 & 0\\
0 & 0 & 0 & 0 &  0 & 0 & 0 & 0\\
0 & 0 & 0 & -\tau_0 &  0 & 0 &  0 & 0\\
0 & 0 & 0 & 0 & -\tau_0 & 0 & 0 & 0\\
0 & 0 & 0 & 0 & 0 & 0 & 0 & 0\\ \hline
0 & 0 & 0 & 0 & 0 & 0 & \omega_0^2 &  0\\
0 & 0 & 0 & 0 & 0 & 0 & 0 & \omega_0^2\\
\end{array}\right]\mymatrix{\delta  x_{O\bar G}^\mathcal{I}\\ \delta  y_{O\bar G}^\mathcal{I}\\ \delta  z_{O\bar G}^\mathcal{I}\\ \delta \theta_{\mathcal{I}\mathcal{B}x}\\ \delta  \theta_{\mathcal{I}\mathcal{B}y}\\ \delta  \theta_{\mathcal{I}\mathcal{B}z}\\ \hline \eta_\theta \\ \eta_\phi} = \mymatrix{\delta F_{\bar G x}^\mathcal{I}\\\delta F_{\bar G y}^\mathcal{I}\\\delta F_{\bar G z}^\mathcal{I}\\\delta \tau_x^\mathcal{B}\\\delta \tau_y^\mathcal{B}\\\delta \tau_z^\mathcal{B}\\\hline 0\\0}
\end{split}\label{linmodal}\end{equation}
\normalsize
where we defined
\begin{align}
    m_{\bar G} &:= m_B + m_P\\
    \omega_0^2 &:= \frac{\bar F_z}{m_{\bar G}\ell}\label{omegazeroquadro}\\
    \tau_0 &:= \frac{m_P}{m_{\bar G}}\bar F_z \ell\\
    I^\mathcal{B} &:= J_{\bar G}^\mathcal{B} - m_P\crossmat{{\bm{r}}_{\bar G Q}^\mathcal{B}}^2
\end{align}
and from which we can clearly identify the modal participation matrix of the pendulum
\begin{equation}
L := \sqrt{m_P}\cdot \mymatrix{
    1 & 0 & 0 & 0 & z_{\bar GQ}^\mathcal{B}  & -y_{\bar GQ}^\mathcal{B}\\
    0 & 1 & 0 & -z_{\bar GQ}^\mathcal{B} & 0 & x_{\bar GQ}^\mathcal{B}
}
\end{equation}
Notice how the vertical translational dynamics correspond to a simple, decoupled, double integrator.

\paragraph{Modal damping}
The damping coefficient $q$, expressed in [kg m/s], can be related to the modal damping ratio $\xi$ through
\begin{equation}
    q = 2\xi m_P\sqrt{\frac{\bar F_z \ell}{m_{\bar G}}}
\end{equation}
which follows directly from the equivalent second-order linearized pendulum dynamics.

\subsection{Single pendulum: force in body frame}
If we choose a nominal body-fixed force $\bar {\bm{F}}_{\bar G}^\mathcal{B} = \bar F_z {\bm{e}}_z$ instead of a nominal inertial force, then
\begin{equation}
    \delta {\bm{F}}_{\bar G}^\mathcal{I} = \bar R_{\mathcal{I}\mathcal{B}} \delta {\bm{F}}_{\bar G}^\mathcal{B} + \delta R_{\mathcal{I}\mathcal{B}} \bar F_z {\bm{e}}_z
\end{equation}
which leads to a slightly different modal stiffness matrix than the one presented in Eq.~\eqref{linmodal}
\begin{equation}
K = \left[\begin{array}{cccccc|cc}
0 & 0 & 0 & 0 & -\bar F_z & 0 & 0 & 0\\
0 & 0 & 0 & \bar F_z & 0 & 0 &  0 & 0\\
0 & 0 & 0 & 0 &  0 & 0 & 0 & 0\\
0 & 0 & 0 & -\tau_0 &  0 & 0 &  0 & 0\\
0 & 0 & 0 & 0 & -\tau_0 & 0 & 0 & 0\\
0 & 0 & 0 & 0 & 0 & 0 & 0 & 0\\ \hline
0 & 0 & 0 & 0 & 0 & 0 & \omega_0^2 &  0\\
0 & 0 & 0 & 0 & 0 & 0 & 0 & \omega_0^2\\
\end{array}\right]
\label{stiffbodyframe}
\end{equation}
The linearized mass matrix is the same as in Eq.~\eqref{linmodal}.

\subsection{Multiple pendulums}
The elements in Eq.~\eqref{linmodal} and Eq.~\eqref{stiffbodyframe} can be \emph{straightforwardly} generalized when additional pendulums are added to the rigid body, as done in Section \ref{secmultipen}, yielding
\begin{align}
    M = \mymatrix{M_{rr} & L_1^\top & \dots & L_n^\top\\ L_1 & I_{2\times 2}\\ \vdots & & \ddots \\ L_n & & & I_{2\times 2}}&& K = \mymatrix{K_{rr} & 0 & \dots & 0\\ 0 &\omega_{0_1}^2 I_{2\times 2}\\ \vdots &  &\ddots\\ 0 & & & \omega_{0_n}^2 I_{2\times 2}}
\end{align}
where we have defined
\begin{align}
    M_{rr} & := \mymatrix{m_{\bar G}I_{3\times 3} & -\sum_i m_{P_i} \ell_i \crossmat{{\bm{e}}_z}\\ \sum_i m_{P_i} \ell_i \crossmat{{\bm{e}}_z} & I^\mathcal{B}}\\
    K_{rr} & := \mymatrix{0_{3\times 3} & \kappa \bar F_z \crossmat{{\bm{e}}_z}\\ 0_{3\times 3} & \sum_i \tau_{0_i} \crossmat{{\bm{e}}_z}^2}\\
    L_i & := \sqrt{m_{P_i}}\cdot \mymatrix{
    1 & 0 & 0 & 0 & z_{\bar GQ_i}^\mathcal{B}  & -y_{\bar GQ_i}^\mathcal{B}\\
    0 & 1 & 0 & -z_{\bar GQ_i}^\mathcal{B} & 0 & x_{\bar GQ_i}^\mathcal{B}\\
    }
\end{align}
and
\begin{align}
    m_{\bar G} &:= m_B + \sum_i m_{P_i}\\
    \omega_{0_i}^2 &:= \frac{\bar F_z}{m_{\bar G}\ell_i}\\
    \tau_{0_i} &:= \frac{m_{P_i}}{m_{\bar G}}\bar F_z \ell_i\\
    I^\mathcal{B} &:= J_{\bar G}^\mathcal{B} - \sum_i m_{P_i}\crossmat{{\bm{r}}_{\bar G Q_i}^\mathcal{B}}^2
\end{align}
with $\kappa = 1$ if we use forces in body frame, and zero otherwise.


\section{Mass-spring-damper model} 
The mass--spring--damper model provides an alternative equivalent mechanical representation of propellant sloshing. This section presents a rigorous derivation establishing the exact equivalence between the linearized pendulum model and the linearized mass--spring--damper dynamics. In doing so, it further highlights how the presence of a nominal longitudinal force modifies the linearized behavior of the coupled system.

For simplicity, a two-dimensional mass--spring system is considered, neglecting damping. The sloshing mass is assumed to be concentrated at point $P$, while linear springs of stiffness $k$ act along the ${\bm{x}}_B$ and ${\bm{y}}_B$ directions. No relative motion between the sloshing mass and the rigid body is allowed along ${\bm{z}}_B$. Under these assumptions, and following the same modeling and linearization procedure adopted for the pendulum formulation with respect to $\bar G$, the nonlinear equations governing the rigid-body dynamics and the mass--spring translational motion can be written as
\begin{align}
m_B \ddot {\bm{r}}_{O\bar G}^\mathcal{I} + m_B R_{\mathcal{I}\mathcal{B}}\crossmat{{\bm{r}}_{B\bar G}^\mathcal{B}} \dot {\bm{\omega}}_{\mathcal{I}\mathcal{B}}^\mathcal{B} - m_B R_{\mathcal{I}\mathcal{B}}\crossmat{{\bm{\omega}}_{\mathcal{I}\mathcal{B}}^\mathcal{B}}^2{\bm{r}}_{B \bar G}^\mathcal{B} &= {\bm{F}}_{\bar G}^\mathcal{I} + m_B {\bm{g}}^\mathcal{I} - R_{\mathcal{I}\mathcal{B}}{\bm{N}}^\mathcal{B}\\
-m_B \crossmat{{\bm{r}}_{B\bar G}^\mathcal{B}} R_{\mathcal{B}\mathcal{I}} \ddot {\bm{r}}_{O\bar G}^\mathcal{I} + J_{\bar G}^\mathcal{B} \dot {\bm{\omega}}_{\mathcal{I}\mathcal{B}}^\mathcal{B} + {\bm{\omega}}_{\mathcal{I}\mathcal{B}}^\mathcal{B} \times J_{\bar G}^\mathcal{B}{\bm{\omega}}_{\mathcal{I}\mathcal{B}}^\mathcal{B} &= {\bm{\tau}}^\mathcal{B} - {\bm{r}}_{B\bar G}^\mathcal{B} \times m_B {\bm{g}}^\mathcal{B} - {\bm{r}}_{\bar GP}^\mathcal{B} \times {\bm{N}}^\mathcal{B}\\
m_P \ddot {\bm{r}}_{OP}^\mathcal{I} &= m_P {\bm{g}}^\mathcal{I} + R_{\mathcal{I}\mathcal{B}}{\bm{N}}^\mathcal{B}\label{masssprnging0}
\end{align}
The force exchanged between the rigid body and the mass--spring can be explicitly written as
\[{\bm{N}}^\mathcal{B} = \mymatrix{-k \left(x_{\bar GP}^\mathcal{B} - x_0^\mathcal{B}\right)\\ -k \left(y_{\bar GP}^\mathcal{B} - y_0^\mathcal{B}\right)\\ N_z}\]
where $x_0^\mathcal{B}$ and $y_0^\mathcal{B}$ are the lateral coordinates of $m_P$ such that no elastic force is generated. $N_z$ is the force exchanged along ${\bm{z}}_B$ between the two bodies that ensures
\begin{equation}
    \ddot z_{\bar GP}^\mathcal{B} = \dot z_{\bar GP}^\mathcal{B} = 0\label{staticmasps}
\end{equation}
at all times, with $z_{\bar GP}^\mathcal{B} = \bar z_{\bar GP}^\mathcal{B}$.
Knowing that, from the kinematics
\[\ddot {\bm{r}}_{OP}^\mathcal{I} = \ddot {\bm{r}}_{O\bar G}^\mathcal{I} - R_{\mathcal{I}\mathcal{B}} \crossmat{{\bm{r}}_{\bar GP}^\mathcal{B}}\dot {\bm{\omega}}_{\mathcal{I}\mathcal{B}}^\mathcal{B} + R_{\mathcal{I}\mathcal{B}} \ddot {\bm{r}}_{\bar GP}^\mathcal{B} + 2 R_{\mathcal{I}\mathcal{B}} \crossmat{{\bm{\omega}}_{\mathcal{I}\mathcal{B}}^\mathcal{B}}\dot {\bm{r}}_{\bar GP}^\mathcal{B} + R_{\mathcal{I}\mathcal{B}} \crossmat{{\bm{\omega}}_{\mathcal{I}\mathcal{B}}^\mathcal{B}}^2{\bm{r}}_{\bar GP}^\mathcal{B}\]
we can rewrite the dynamics of the mass-spring in Eq.~\eqref{masssprnging0} as
\begin{equation}
m_P R_{\mathcal{B}\mathcal{I}}\ddot {\bm{r}}_{O\bar G}^\mathcal{I} - m_P {{\bm{r}}_{\bar GP}^\mathcal{B}}\times \dot {\bm{\omega}}_{\mathcal{I}\mathcal{B}}^\mathcal{B} + m_P \ddot {\bm{r}}_{\bar GP}^\mathcal{B} = m_P {\bm{g}}^\mathcal{B} + {\bm{N}}^\mathcal{B} - m_P \crossmat{{\bm{\omega}}_{\mathcal{I}\mathcal{B}}^\mathcal{B}} \left(2 \dot {\bm{r}}_{\bar GP}^\mathcal{B} + \crossmat{{\bm{\omega}}_{\mathcal{I}\mathcal{B}}^\mathcal{B}}{\bm{r}}_{\bar GP}^\mathcal{B} \right) \label{masssprnging02}
\end{equation}
from which we can get the expression of the nominal exchanged force
\begin{equation}
    \bar {\bm{N}}^\mathcal{B} = m_P \ddot {\bar{\bm{r}}}_{O\bar G}^\mathcal{I} - m_P{\bm{g}}^\mathcal{I}
\label{msdN}\end{equation}
and it's first order variation, which also serves as linearization of Eq.~\eqref{masssprnging02} along the nominal trajectory
\begin{equation}
    \delta {\bm{N}}^\mathcal{B} = m_P \delta \ddot {\bm{r}}_{O\bar G}^\mathcal{I} + \crossmat{\bar {\bm{N}}^\mathcal{B}}\delta {\bm{\theta}}_{\mathcal{I}\mathcal{B}} - m_P \crossmat{\bar {\bm{r}}_{\bar GP}^\mathcal{B}}\delta \dot {\bm{\omega}}_{\mathcal{I}\mathcal{B}}^\mathcal{B} + m_P \delta \ddot {\bm{r}}_{\bar GP}^\mathcal{B}
    \label{msddN}
\end{equation}
Using Eqs.~\eqref{comrela}, \eqref{msdN}, and \eqref{msddN}, we can obtain the nominal position and attitude dynamics as
\begin{align}
    m_{\bar G} \ddot {\bar{\bm{r}}}_{O\bar G}^\mathcal{I} & = \bar F_z {\bm{e}}_z + m_{\bar G} {\bm{g}}^\mathcal{I} \label{eqtramsmd}\\
    \bar {\bm{\tau}}^\mathcal{B} & = 0
\end{align}
as well as their linearized versions
\begin{align}
    m_{\bar G} \delta \ddot {\bm{r}}_{O\bar G}^\mathcal{I} + m_P \delta \ddot {\bm{r}}_{\bar GP}^\mathcal{B} = \delta {\bm{F}}_{\bar G}^\mathcal{I} \label{msdlinpps}\\
 \left(J_{\bar G}^\mathcal{B} - m_P \crossmat{\bar {\bm{r}}_{\bar GP}^\mathcal{B}}^2\right) \delta \dot {\bm{\omega}}_{\mathcal{I}\mathcal{B}}^\mathcal{B} + m_P {\bar {\bm{r}}_{\bar GP}^\mathcal{B}}\times\delta \ddot {\bm{r}}_{\bar GP}^\mathcal{B} = \delta {\bm{\tau}}^\mathcal{B} + {\bar {\bm{N}}^\mathcal{B}}\times \delta {\bm{r}}_{\bar GP}^\mathcal{B}\label{msdlinatt}
\end{align}
Combining Eq.~\eqref{eqtramsmd} and \eqref{msdN} we can obtain an expression of the nominal exchanged force
\begin{equation}
    \bar {\bm{N}}^\mathcal{B} = \frac{m_P}{m_{\bar G}} \bar F_z {\bm{e}}_z = \bar N_z {\bm{e}}_z
\end{equation}
and therefore demonstrate that $x_0^\mathcal{B} = \bar x_{\bar GP}^\mathcal{B}$, $y_0^\mathcal{B} = \bar y_{\bar GP}^\mathcal{B}$. We can now regroup Eqs.~\eqref{msdlinpps}, \eqref{msdlinatt}, and \eqref{msddN} into the following system of equations
\[\begin{split}
&\mymatrix{m_{\bar G} I_{3\times 3} & 0 & m_P I_{3\times 3}\\
0 & J_{\bar G}^\mathcal{B} -m_P \crossmat{\bar{\bm{r}}_{\bar GP}^\mathcal{B}}^2& m_P \crossmat{\bar{\bm{r}}_{\bar GP}^\mathcal{B}}\\
m_P I_{3\times 3} & - m_P \crossmat{\bar {\bm{r}}_{\bar GP}^\mathcal{B}} & m_P I_{3\times 3}}\mymatrix{\delta \ddot {\bm{r}}_{O\bar G}^\mathcal{I} \\\delta \dot {\bm{\omega}}_{\mathcal{I}\mathcal{B}}^\mathcal{B}\\\delta \ddot {\bm{r}}_{\bar GP}^\mathcal{B}} + \\ 
& \hspace{3cm} + \mymatrix{0 & 0 & 0\\
0 & 0 & -\crossmat{\bar {\bm{N}}^\mathcal{B}}\\ 0 & \crossmat{\bar {\bm{N}}^\mathcal{B}} & -k \crossmat{{\bm{e}}_z}^2}\mymatrix{\delta {\bm{r}}_{O\bar G}^\mathcal{I} \\ \delta {\bm{\theta}}_{\mathcal{I}\mathcal{B}}\\ \delta {\bm{r}}_{\bar GP}^\mathcal{B}} = \mymatrix{\delta {\bm{F}}_{\bar G}^\mathcal{I}\\\delta {\bm{\tau}}^\mathcal{B}\\0\\0\\\delta N_z}
\end{split}\]
Dropping the terms in $\delta \ddot z_{\bar GP}^\mathcal{B}$ and $\delta z_{\bar GP}^\mathcal{B}$, which, by hypothesis, are both always equal to zero (no relative movement between $m_B$ and $m_P$ is allowed along ${\bm{z}}_B$), and introducing the following modal coordinates
\begin{align}
    \eta_x & := \sqrt{m_P}\left(\delta x_{\bar GP}^\mathcal{B} - \frac{\bar N_z}{k} \delta \theta_{\mathcal{I}\mathcal{B}y}\right)\\
    \eta_y & := \sqrt{m_P}\left(\delta y_{\bar GP}^\mathcal{B} + \frac{\bar N_z}{k} \delta \theta_{\mathcal{I}\mathcal{B}x}\right)
\end{align}
we can transform the system above using the same approach implemented with the pendulum, i.e., coordinate transformation, algebraic manipulation, and scaling the last two equations by $\sqrt{m_P}$. If we set $k = m_P \omega_0^2$ with $\omega_0^2$ as in Eq.~\eqref{omegazeroquadro}, we obtain the same exact representation already preseted in Eq.~\eqref{linmodal}, demonstrating that the linearized dynamics of a mass--spring--damper is perfectly equivalent to the one of a 3D pendulum.

\paragraph{What if there is no nominal force?} 
If $\bar F_z = 0$, and still assuming $k > 0$, the linearized equations of motion become what one could "expect" for this system
\[\begin{split}
\left[\begin{array}{cccccc|cc}
m_{\bar G}  & 0 & 0 & 0 & 0 & 0 & \sqrt{m_P}  & 0\\
0 & m_{\bar G} & 0 &  0 & 0 & 0 & 0 & \sqrt{m_P}\\
0 & 0 &  m_{\bar G}& 0 & 0  & 0 &  0 & 0\\
0 & 0 &  0 & I_{xx}^\mathcal{B}  & I_{xy}^\mathcal{B} &  I_{xz}^\mathcal{B} & 0 & -\sqrt{m_P}z_{\bar GP}^\mathcal{B}\\
0 & 0 & 0 & I_{xy}^\mathcal{B}  & I_{yy}^\mathcal{B} &  I_{yz}^\mathcal{B} & \sqrt{m_P} z_{\bar GP}^\mathcal{B} & 0\\
0 & 0  & 0 & I_{xz}^\mathcal{B} & I_{yz}^\mathcal{B} &  I_{zz}^\mathcal{B}  & -\sqrt{m_P} y_{\bar GP}^\mathcal{B} &  \sqrt{m_P} x_{\bar GP}^\mathcal{B}\\ \hline
\sqrt{m_P} & 0 & 0 & 0 & \sqrt{m_P}z_{\bar GP}^\mathcal{B}  & -\sqrt{m_P} y_{\bar GP}^\mathcal{B}  & 1 & 0\\
0 & \sqrt{m_P} & 0 & -\sqrt{m_P} z_{\bar GP}^\mathcal{B} & 0 & \sqrt{m_P} x_{\bar GP}^\mathcal{B} & 0 & 1
\end{array}\right]\mymatrix{\delta  \ddot x_{O\bar G}^\mathcal{I}\\ \delta  \ddot y_{O\bar G}^\mathcal{I}\\ \delta \ddot  z_{O\bar G}^\mathcal{I}\\ \delta \ddot \theta_{\mathcal{I}\mathcal{B}x}\\ \delta \ddot  \theta_{\mathcal{I}\mathcal{B}y}\\ \delta \ddot  \theta_{\mathcal{I}\mathcal{B}z}\\ \hline \ddot \eta_x \\ \ddot \eta_y} + \\
+ 
\left[\begin{array}{cccccc|cc}
0 & 0 & 0 & 0 & 0 & 0 & 0 & 0\\
0 & 0 & 0 & 0 & 0 & 0 &  0 & 0\\
0 & 0 & 0 & 0 &  0 & 0 & 0 & 0\\
0 & 0 & 0 & 0 &  0 & 0 &  0 & 0\\
0 & 0 & 0 & 0 & 0 & 0 & 0 & 0\\
0 & 0 & 0 & 0 & 0 & 0 & 0 & 0\\ \hline
0 & 0 & 0 & 0 & 0 & 0 & \omega_0^2 &  0\\
0 & 0 & 0 & 0 & 0 & 0 & 0 & \omega_0^2\\
\end{array}\right]\mymatrix{\delta  x_{O\bar G}^\mathcal{I}\\ \delta  y_{O\bar G}^\mathcal{I}\\ \delta  z_{O\bar G}^\mathcal{I}\\ \delta \theta_{\mathcal{I}\mathcal{B}x}\\ \delta  \theta_{\mathcal{I}\mathcal{B}y}\\ \delta  \theta_{\mathcal{I}\mathcal{B}z}\\ \hline \eta_x \\ \eta_y} = \mymatrix{\delta F_{\bar G x}^\mathcal{I}\\\delta F_{\bar G y}^\mathcal{I}\\\delta F_{\bar G z}^\mathcal{I}\\\delta \tau_x^\mathcal{B}\\\delta \tau_y^\mathcal{B}\\\delta \tau_z^\mathcal{B}\\\hline 0\\0}
\end{split}\]
where $I^\mathcal{B} := J_{\bar G}^\mathcal{B} -m_P \crossmat{\bar{\bm{r}}_{\bar GP}^\mathcal{B}}^2$. It is obvious that having $\bar F_z = 0$ and $k > 0$ is a meaningless working hypothesis for the pendulum, as the pendulum needs a longitudinal force to\dots oscillate like a pendulum!

The $\bar F_z = 0$ case can also serve as a simple 0-g sloshing model. If we add a third modal degree of freedom $\eta_z$ (as is needed for a 0-g model), the modal participation matrix becomes
\[L_{\text{0-g}} = \sqrt{m_P}\cdot \mymatrix{I_{3\times 3} & -\crossmat{\bar{\bm{r}}_{\bar GP}^\mathcal{B}}}\]

\section{Model validation}
\begin{figure}
    \centering
    \includegraphics[width=0.9\textwidth]{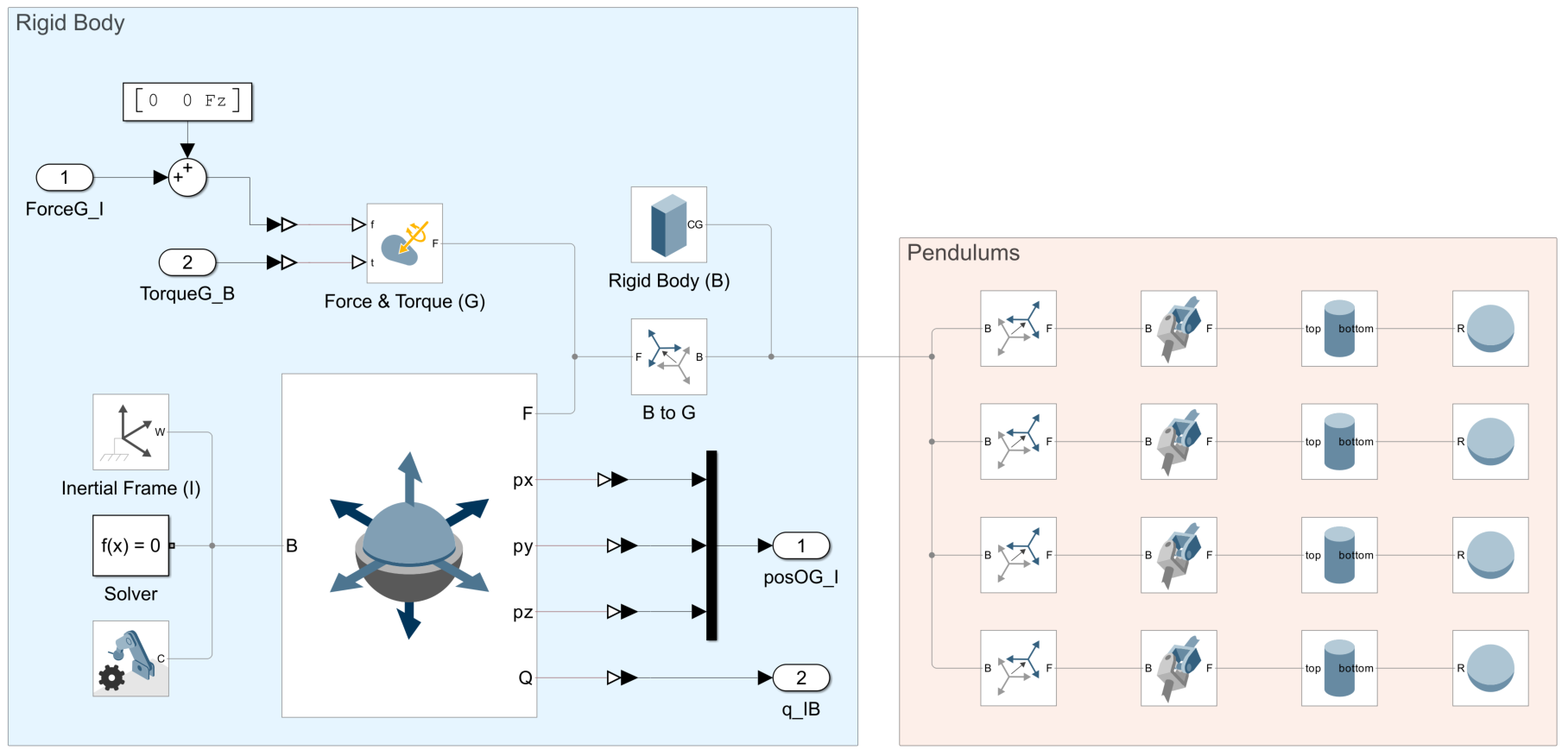}
    \caption{MATLAB Simscape validation model}
    \label{simscape}
\end{figure}

Validation of both the nonlinear and linearized models was performed through comparison with a MATLAB Simscape Multibody implementation \cite{b3}. The validation model consisted of a multibody system comprising a rigid hub and four pendulums, as illustrated in Figure~\ref{simscape}.

\subsection{Nonlinear model}
\begin{figure}
    \centering
    \includegraphics[width=\textwidth]{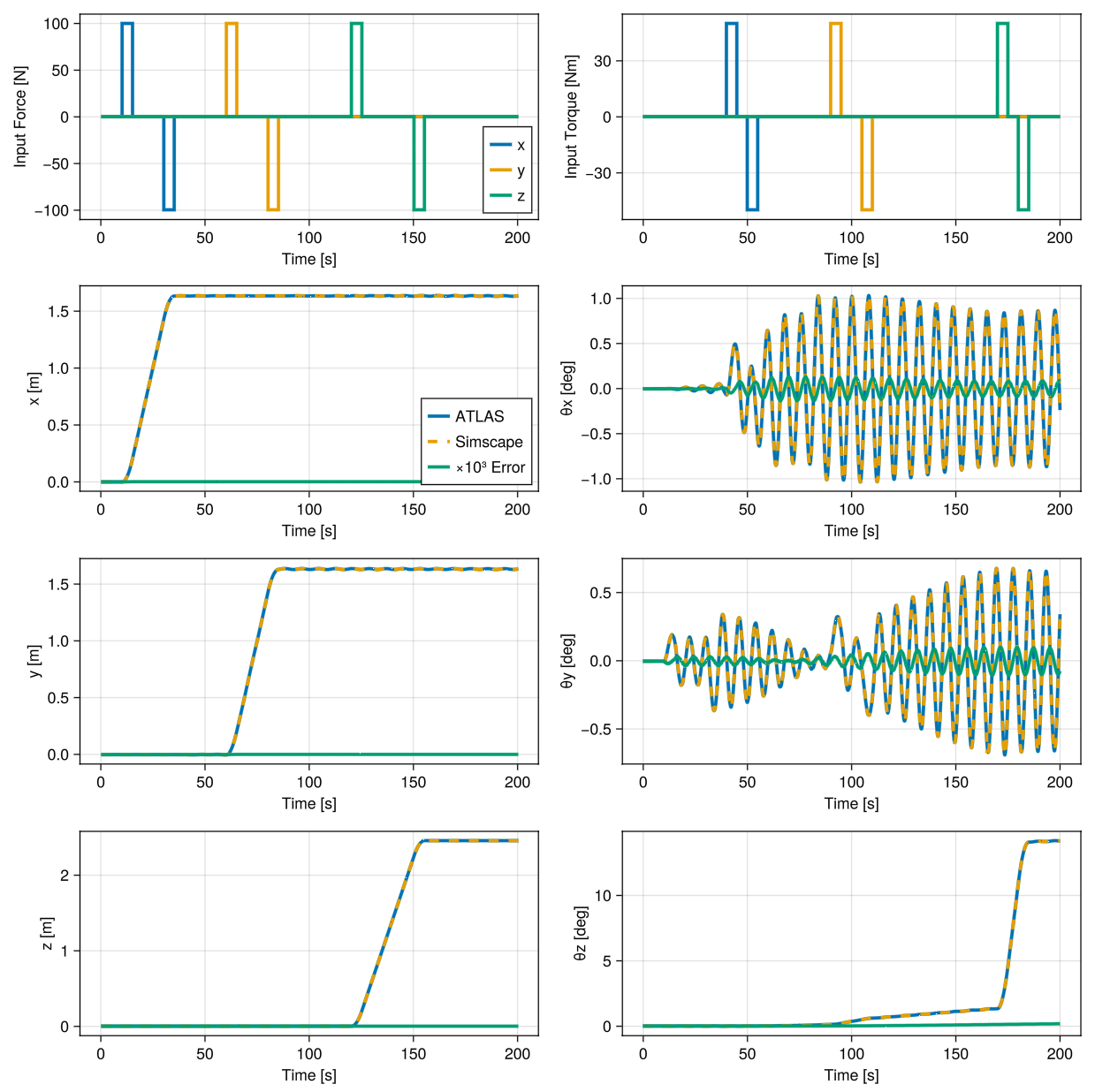}
    \caption{Time response of the nonlinear models to open-loop inputs.}
    \label{nlval}
\end{figure}

The nonlinear model was validated by applying open-loop force and torque inputs to the system and comparing its time response against that of the corresponding Simscape model. The applied input profiles and the resulting responses are shown in Figure~\ref{nlval}, together with the difference between the Simscape and analytical model responses, shown in green and scaled by a factor of 1000 for visibility. The results demonstrate an excellent agreement between the two models. The small residual discrepancies are attributed to differences in the numerical integration schemes used for the simulations. Specifically, Simscape employs its own fixed-step fifth-order Runge--Kutta integrator, whereas the analytical model was implemented in Julia using the fixed-step \texttt{DP5()} integrator from \emph{DifferentialEquations.jl}. In both cases, a timestep of 1~ms was adopted. The maximum error over the entire simulation remains below 0.002\% of the maximum absolute value of the corresponding response for all position and attitude states.

\subsection{Linear model}
\begin{figure}
    \centering
    \includegraphics[width=\textwidth]{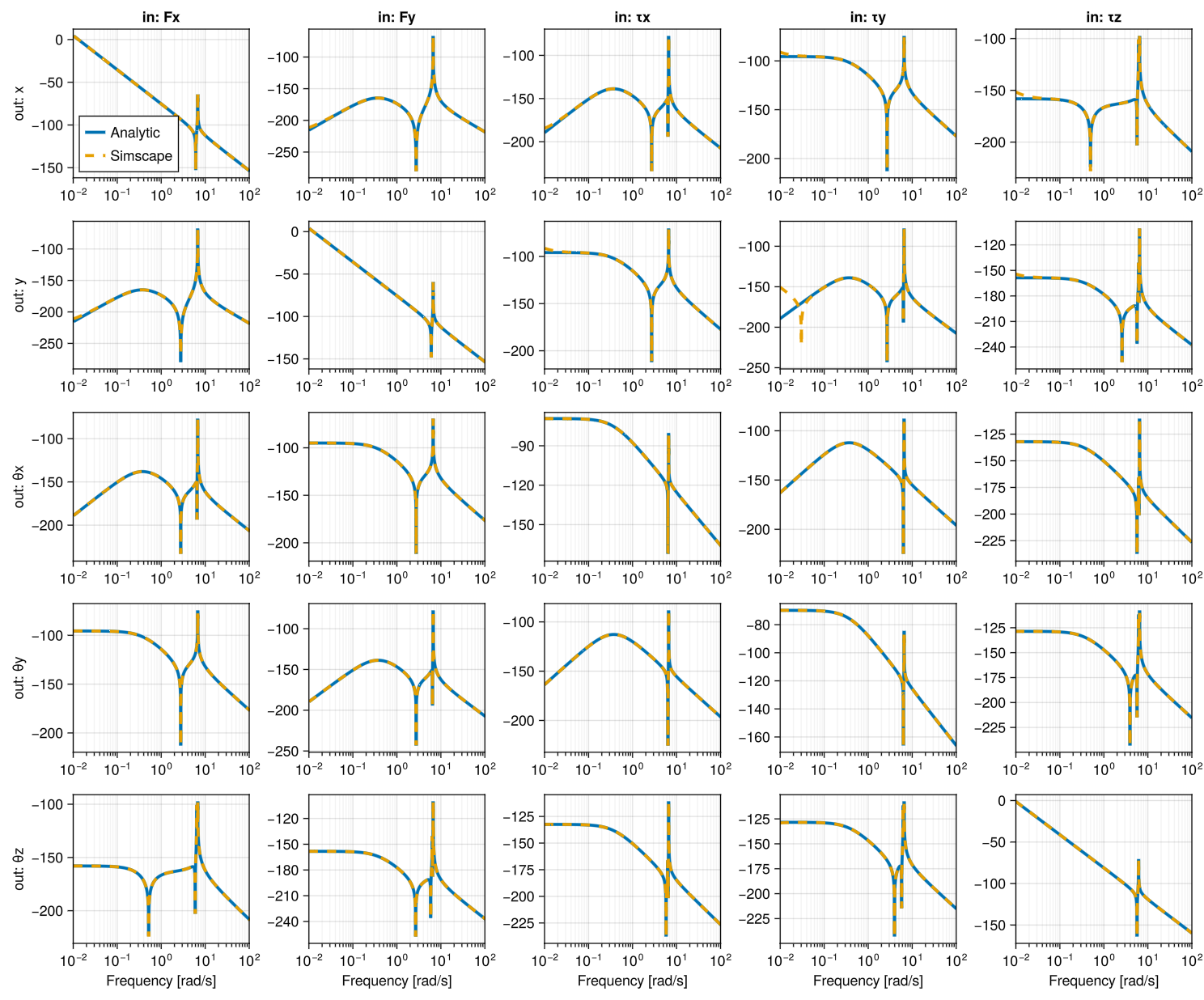}
    \caption{Bode magnitude plot [dB] of the linearized model, forces in inertial frame.}
    \label{linvalI}
\end{figure}

\begin{figure}
    \centering
    \includegraphics[width=\textwidth]{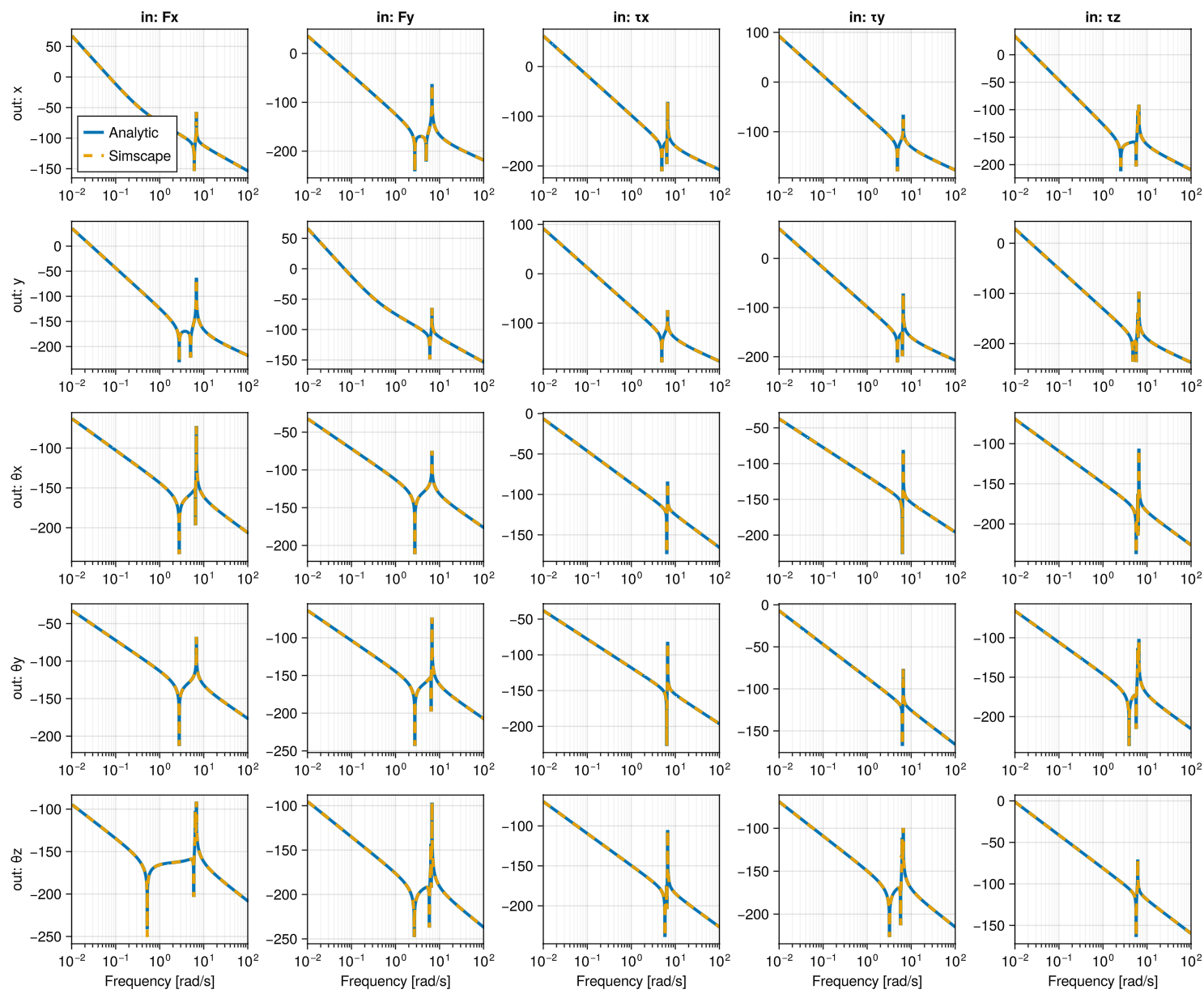}
    \caption{Bode magnitude plot [dB] of the linearized model, forces in body frame.}
    \label{linvalB}
\end{figure}

Validation of the linearized models was performed by comparing the Bode plots of the analytical model with those obtained from the linearized Simscape plant. The comparison was carried out considering force inputs expressed both in the inertial frame (Figure~\ref{linvalI}) and in the body-fixed frame (Figure~\ref{linvalB}). The vertical translational dynamics, which correspond to a decoupled double integrator, are omitted from the plots. The results show an excellent agreement between the two models across the entire frequency range. A minor discrepancy can be observed at low frequencies when forces are expressed in the inertial frame. Since no physical mechanism can account for this behavior, it is attributed to the finite numerical accuracy of the Simscape linearization process.

\section{Conclusion}
This work presented both nonlinear and linearized equivalent mechanical models for the analysis of propellant sloshing under high-g conditions. Furthermore, the equivalence between the linearized pendulum formulation and the classical mass--spring--damper representation was formally established. The proposed models were validated through comparisons with corresponding MATLAB Simscape implementations in both the time and frequency domains. The excellent agreement obtained in all test cases confirms the correctness of the derived formulations.


\begin{thebibliography}{00}
\bibitem{b1} F. T. Dodge, ``The New "Dynamic Behavior of Liquids in Moving Containers"'', Southwest Research Institute, 2000.
\bibitem{b2} P. Cappuccio, C. Allard, H. Schaub, ``Fully-Coupled Spherical Modular Pendulum Model to Simulate Spacecraft Propellant Slosh'', AAS/AIAA Astrodynamics Specialist Conference, 2018.
\bibitem{b3} Simscape Multibody product webpage: https://www.mathworks.com/products/simscape-multibody.html, accessed 13/11/2025.
\end{thebibliography}
\end{document}